# Negative ion radio frequency surface plasma source with solenoidal magnetic field

V. Dudnikov[1,a], R.P. Johnson[1], B. Han[2], Y. Kang[2], S. Murray[2], T. Pennisi[2], C. Piller[2], C. Stinson[2], M. Stockli[2], R. Welton[2], G. Dudnikova[3]

[1]*Muons, Inc., Batavia, IL 60510, USA,;* [2]*ORNL, Oak Ridge, TN 37831, USA;;* [3]*Institute of Computational Technologies SBRAS, Novosibirsk, Russia*

[a]*Corresponding Author:Vadim@muonsinc.com*

Pulsed and CW operation of negative ion radio frequency surface plasma source with a solenoidal magnetic field is described. Dependences of a beam current on RF power, extraction voltage, solenoid magnetic field, gas flow are presented. Compact design of RF SPS is presented.

## INTRODUCTION

Efficiency of plasma generation in a Radio Frequency (RF) ion source can be increased by application of a solenoidal magnetic field [1-9]. The specific efficiency of positive ion generation was improved by the solenoidal magnetic field, from 5 mA/cm$^2$ kW to 200 mA/cm$^2$ kW [8,9]. Chen [10] presented an explanation for the concentration of plasma density near the axis by a magnetic field through a short circuit in the plasma plate. Additional concentration factor can be a secondary ion-electron emission initiated by high positive potential of plasma relative the plasma plate. Secondary negative ion emission can be increased by cesiation-injection of cesium [11-14], increasing a secondary electron and photo emission.

## RF ION SOUIRCE IN SNS TEST STAND

An RF SPS was installed at the Spallation Neutron Source (SNS) test stand. A design of ion source and Low Energy Beam Transportation channel (LEBT) is shown in Fig. 1[15]. The RF ion source consists of an AlN ceramic chamber with a cooling jacket from keep. At the left side, an RF assisted triggering plasma gun (TPG) is attached. At the right side, a plasma electrode with an extraction system is attached. The discharge chamber is surrounded by a saddle (or solenoidal) antenna. A solenoid is located around the antenna. The LEBT at the right side consists of an accelerator electrode and two electrostatic lenses which focus a beam into a 7.5 mm diameter hole in the chopper target. The second lens consists of four electrically insulated quadrants, which allows for the chopping of the beam to form an extraction gap inside the accumulator ring. In addition, the voltages on the quadrants can be varied individually to steer the beam in order to improve transmission through the RFQ [15]. A more detailed picture of the extraction system is presented in Fig. 2 a. The plasma plate (1) has a collar (2) with a conical converter surface.

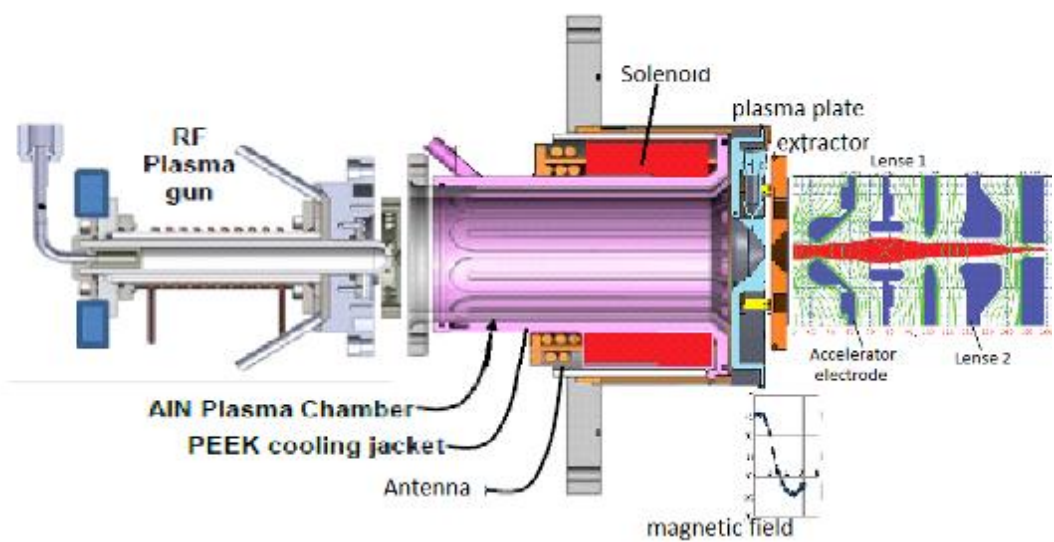

**FIGURE 1:** Design of RF SPS with solenoidal magnetic field and LEBT.

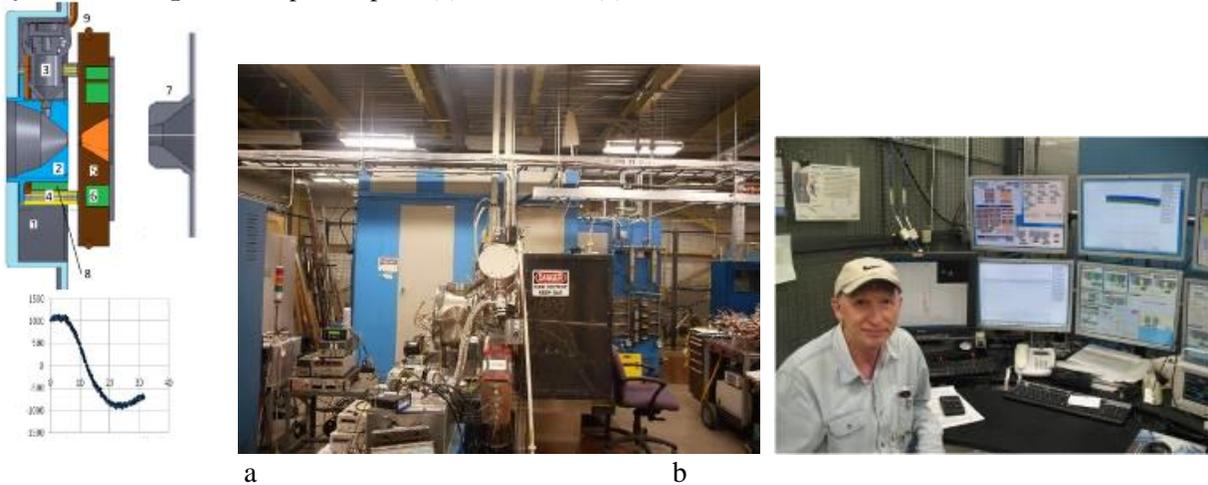

a  b

**FIGURE 2 a-** Design of extraction system.
1- plasma electrode (PE), 2- conical collar with emission aperture, 3- Cs oven, 4- ceramic insulators, 5-extractor electrode, 6- permanent bar magnets, 7- accelerating electrode, 8-ferromagnetic insert, 9- water cooling lines.
**b-**: Large SNS test stand and control system.

Cs oven (3) for the decomposition of $Cs_2CrO_4$ cartridges is attached to the collar. The extractor electrode (5) is attached to the plasma plate through a ceramic insulator (4). Permanent magnet bars (6) are inserted into the water cooled extractor electrode (5). A ferromagnetic insert (8) can be used to shape the magnetic field. The accelerating electrode (7) is used for the acceleration of extracted ions up to 65 keV.

A plasma is generated by a RF current in the antenna. A solenoidal magnetic field concentrates the plasma on the axis. A transverse magnetic field, generated by permanent magnets (6) located inside the water cooled extractor (5), bends the plasma flux and prevents electrons to escape the plasma. A configuration of the transverse magnetic field is shown in Fig. 2 a (below). Any plasma outside a 6 mm diameter circle impacts on the conical surface of a Mo converter, where H- ions are formed. H- production is enhanced by lowering the Mo work function, which is done by adding a partial monolayer of Cs. After plasma conditioning the source for ~3 hours, the Cs chromate is heated to 550 C, which releases Cs from the $Cs_2CrO_4$ cartridges. Negative ions that drift into the source outlet are extracted by the potential of the extractor-electron dump. The extractor, which has a 6 mm aperture, can have a potential up to 8 kV. A 1 kG dipole field integrated into the extractor drives the co-extracted electrons sideways. Most of them are intercepted by the electron dump, which is kept near -57 kV with a +8 kV supply located on the -65 kV platform. The 65-keV H- beam emerges from the extractor and is focused by two electrostatic lenses into the 7.5 mm diameter hole in a chopper target.

The Saddle antenna RF Surface Plasma Source (SA RF SPS) has been prepared for long term testing in the large SNS Test stand. A photograph of the test setup and control system are shown in Fig. 2 b.

## EXPERIMENTAL RESULTS

Typically, 200 W from a 600-W, 13-MHz amplifier generates a TPG continuous low-power plasma. Cathode TPG is biased at 210 V with a current of ~10 mA. The high current beam pulses for 1 ms, up to 60 Hz are generated by up to 50 kW of power from a pulsed 120-kW, 2-Mz amplifier, connected to the antenna through an insulating transformer and matching network. A discharge is triggered at pulsed power Prf= 3.8 kW, antenna current Iant=120 A, antenna voltage Uant=6.5 kV (at 13.56 MHz discharge start from Prf-0.5 kW, Iant=14 A, Uant=1.2kV).

Cesiation is started after 3 hours of conditioning by discharge. The Cs oven is slowly heated up to 550 C. Faraday cup current is increasing in pulse. Chopper target current is increasing and current of the 65 kV power supply increases. The optical spectrum is transformed as shown in Fig. 3.

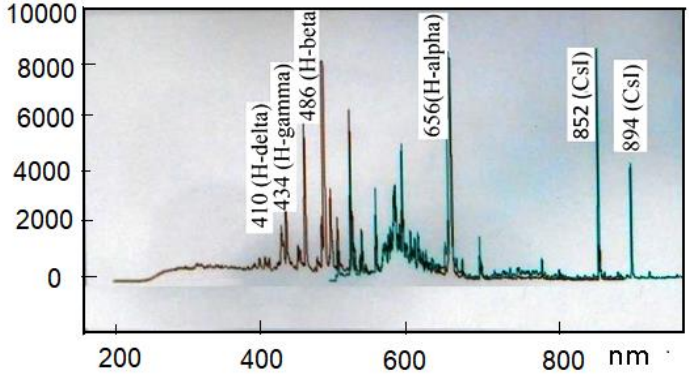

**FIGURE 3.** Optical spectrum of plasma during cesiation.

Cs 852 nm and Cs 894 nm line are increase and new lines are increases. Electrodes of the extraction system and LEBT shine as shown in Fig. 4, but electrode current does not increase significantly.

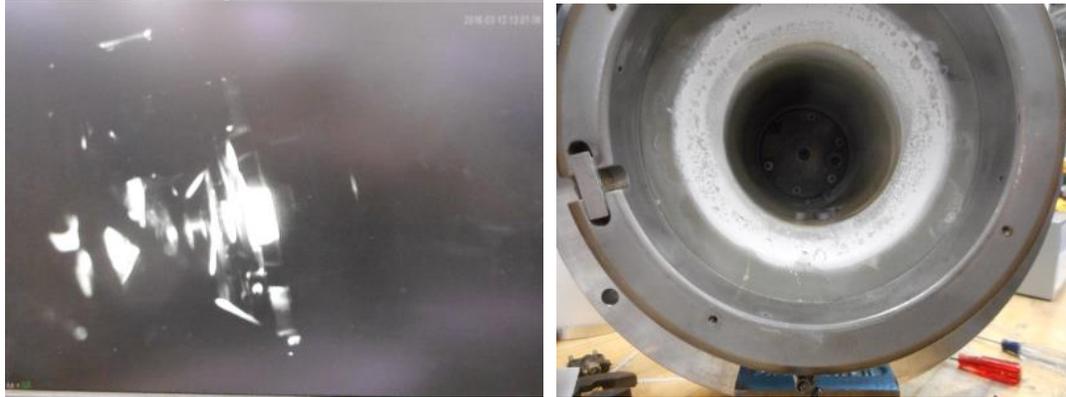

**FIGURE 4.** Electrodes of the extraction system and LEBT is shined during cesiation.
**FIGURE 5**. CsH film on the wall of the discharge chamber.

After cesiation, the extractor current decreases down to 4 mA from 120` mA at RF power Prf=50 kW. Faraday cup current Ifc increases up to 25 mA at Prf=50 kW from the RF generator (9 kW in plasma).

After opening an ion source, a white film of CsH was observed on the discharge chamber wall, not bombarded by the plasma as shown in Fig. 5.

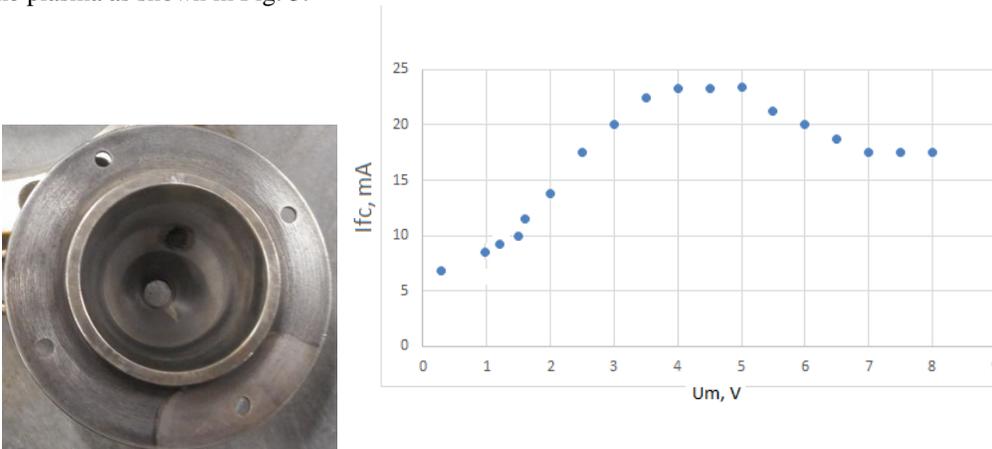

**FIGURE 6.** A dark deposition around the emission aperture of the converter cone.
**FIGURE 7.** Dependents of Faraday cup current Ifc on the magnetic solenoid voltage Um with solenoidal antenna Bt=1000 Gauss. RF power 50 kW from generator (6 kW in the plasma, Um=7 V is correspondent to Bs=250 G).

On the surface of the conical converter, a black plasma deposition around the emission aperture was observed as shown in Fig. 6. Dependence of the Faraday cup current Ifc on magnetic solenoid voltage Um (solenoidal magnetic field Bs) with a solenoidal antenna and Bt=1000 gauss is shown in Fig. 7. With a solenoidal antenna, Ifc increases up to 4 V and decreases after 5 V, because plasma is concentrated up to the emission aperture. With a saddle antenna and Bt=600 Gauss, Ifc increases from 4 mA to 24 mA with increase of Um from 0 to 6 V, as shown in Fig. 8. Dependence of average beam current Ifc on hydrogen flow Q in sccm is shown in Fig. 9. Ifc increases with a decrease of Q.

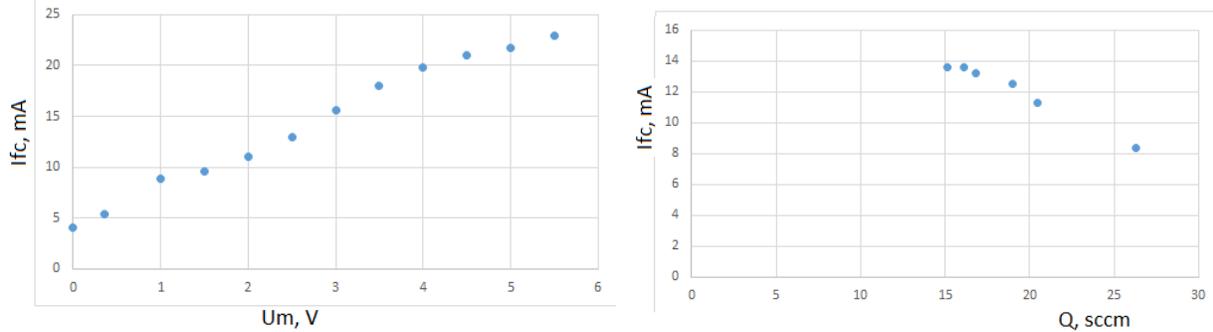

**FIGURE 8.** Dependence of the Faraday cup current Ifc on the magnetic solenoid voltage Um with a saddle antenna. An RF power of 21.5 kW comes from the generator (4 kW in the plasma, Um=7 V is correspondent to Bs=250 G).
**FIGURE 9**. Dependence of average beam current Ifc on hydrogen flow Q in sccm.

## ESTIMATION EFFICIENCY OF H- BEAM GENERATION

Forwarded RF power from the RF generator is measured by a directional coupler and calculated by the following formula:
Prf=42.7 x <I>$^2$ kW, where <I> rms is rms current in V.
Before triggering discharge, all power is dissipated in the insulating transformer, antenna, solenoid and matching network. For our case it is <I>=0.293 V, 3.66 kW, antenna current <I>ant= 83.3 A, antenna voltage V=6,480 V. Active resistance of network + antenna is R=2P/<I>$^2$ant =2*3660/(83.3)$^2$=1 Ohm. For discharge with <I> =0.599 V the power Prf=15.3 kW is dissipated in discharge Pd, in antenna+network Pant and in surrounding antenna solenoid Psol: Prf=Pd+Pant+Psol. For <I>ant1=136 A Pant=R<I>$^2$ant/2=10 kW. Pd=15.3-10=5.3 kW. For Faraday current Ifc=17 mA, the efficiency of current generation is h=3 mA/kW at Um=2.11 V.
At <I>=0.872 V, Prf=34 kW. <I>ant =194.4 A. Pant=20 kW. Ifc=16 mA, h=16/14=1.14 mA/kW at Um=0.
At <I>=0.963 V, Prf=41.7 kW. <I>ant =250 A. Pant=34.3 kW. Pd=41.7-34.3=7.4 kW. Ifc=25 mA, h=25/7.4=3.37 mA/kW at Um=3.2 V.
Volume of the collar is 29 cm$^3$. Mass of the collar is 290 g. A specific thermal permeability of Mo is C=0.255 J/g K.

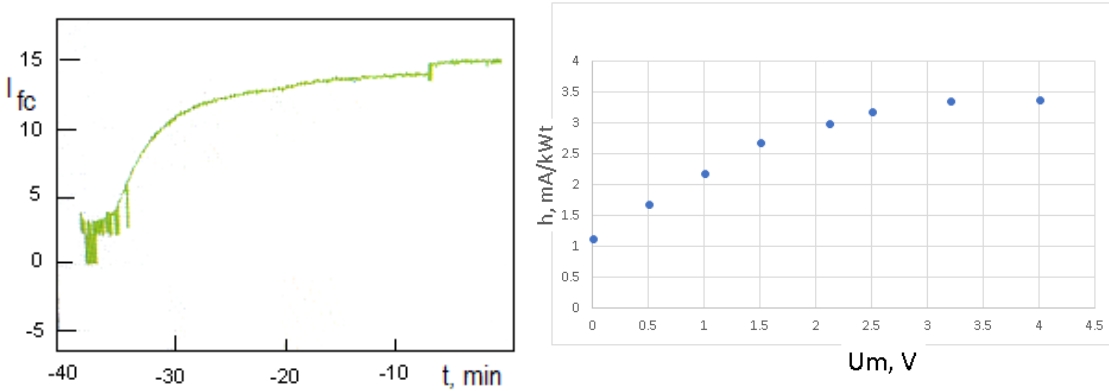

**FIGURE 10**. Cesiation: increase of Faraday cup current (mA) in time during cesiation from 3 mA to 14 mA at constant RF power 40% (9 kW in plasma).
**FIGURE 11**. Dependence of efficiency H- production h on solenoid voltage in mA/kW for pulsed mode operation at 2 MHz.

Thermal permeability of collar is 75 J/C. A speed of the collar cooling after switching off the discharge is 0.7°C/s. Power loss from the collar is 52 W (pulsed power 868 W from Prf=34.2 kW from RF generator (14 kW in plasma) at Um=1.68 V). Slow cesiation: increase of Faraday cup current (mA) in time during cesiation from 3 mA to 14 mA at constant RF power 40% (9 kW in plasma) is shown in Fig. 10. The efficiency of plasma generation and stability of the discharge can be increased by control of the plasma potential [16]. Dependence of efficiency H- production h (in mA/kW) on solenoid voltage for pulsed mode operation at 2 MHz is shown in Fig. 11. It can reach h=3,4 mA/kWt at Um=4 V. Dependence of electron current Ie and collector current IH- on solenoid current Is are shown in Fig 12. At low Is, $I_{H-}$ is ~8 mA and Ie is ~50 mA. With increase of solenoid current the electron current is decreases up to 40 mA, but collector current is increases. After increase solenoid current above 15 A electron current began to increase until 60 mA at solenoid current of 26 A. Collector current is increases up to 20 mA.

## RF ION SOUIRCE IN SMALL SNS TEST STAND

Figure 13 is a schematic of the large RF ion source, showing the AlN ceramic discharge chamber, saddle antenna, and DC solenoid 50 turn, up to 70 A. The chamber has an ID=68 mm. The saddle antenna with inductance L=3.5 μH is made from a water-cooled copper tube. The RF assisted triggering plasma gun (TPG) is attached to the discharge chamber on the left [4,7]. The extraction system is attached on the right side. The plasma in the TPG is generated by a continuous wave (CW) RF discharge (13.56 MHz, ~250 W) and electrons are injected into main discharge chamber by the extraction voltage. The ion source is inserted into a vacuum chamber pumped by a turbo molecular pump. The upper window is used for beam extraction observation. The front window (at right) is used to observe the back side of the collector while it is heated by ion beams.

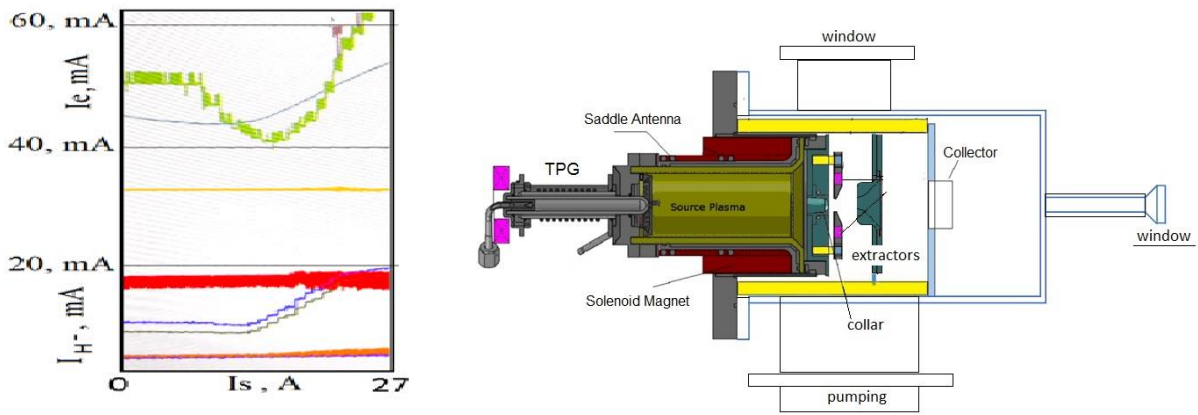

**FIGURE 12.** Evolution of electron current Ie and collector current $I_{H-}$ with increase of solenoid current Is.
**FIGURE 13**. A schematic of the SA ion source with an extraction system, collector in vacuum chamber.

The plasma flux, generated in the AlN discharge chamber by the saddle antenna is guided by the longitudinal magnetic field (created by solenoid) to the plasma electrode, with a conical collar that defines the emission aperture. Ions are extracted from the cone by the extraction voltage Uex between the cone and the extractor attached to the plasma electrode through ceramic insulators. and accelerated by the voltage across the second gap by voltage Uc on the accelerator electrode. A plasma flux is compressed by increase of solenoidal magnetic field.

The evidence of this plasma flux behavior is the trace of the dark film deposited on the conical collar surface. With the increase of the solenoid current, this deposited area decreased up to the emission aperture. The SA ion source was tested with 6 mm diameter emission and extractor apertures. At first was tested a positive ions extraction.

A photograph of proton beam extraction is shown in Fig. 14. It is visible light from positive ion beam and heating of collector up to high temperature. Oscilloscope trace of a collector current is shown in Fig. 15.

Dependence of collector current Ic on extraction voltage is shown in Fig. 16. It is possible to have Ic~50 mA at extraction voltage 16 kV.

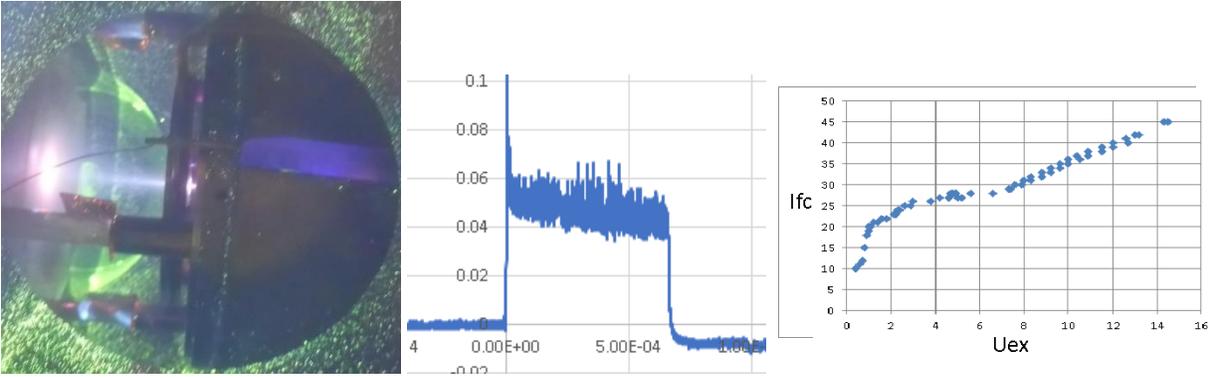

**FIGURE 14.** Photograph of Formation of positive ion beam. Collector is heated up to high temperature
**FIGURE 15**. Signals of positive ion on collector—Ic = 50 mA at RF power ~1.5 kW in the plasma. Time scale is 0.5 ms/div
**FIGURE 16.** Dependence of collector current of positive ions Ic (mA) on extraction voltage (kV), Prf=1.5 kW, Um=3 V.

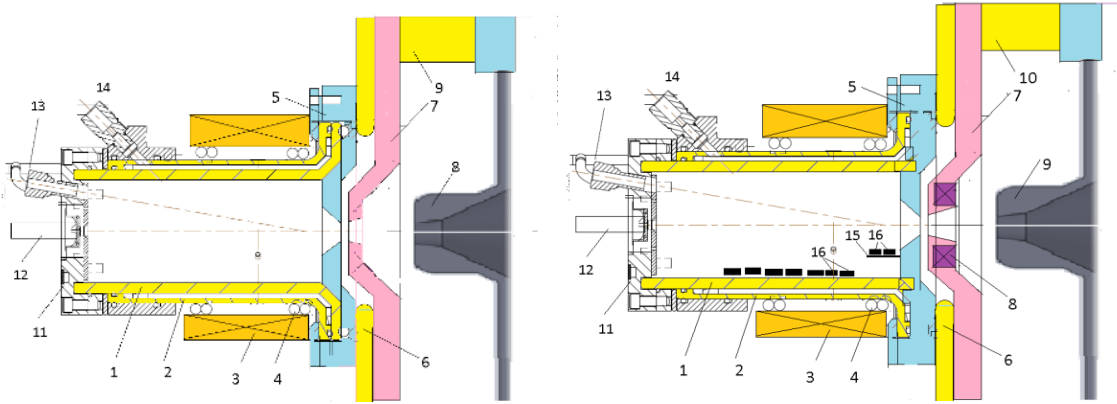

**FIGURE 17.** Schematic of simplified version of RF discharge LV proton source with solenoidal magnetic field.
1. Gas discharge chamber (AlN), 2- cooling jacket from keep, 3- solenoid, 4- helicon antenna, 5-plasma electrode with conical collar and emission aperture, 6-extractor insulator, 7- extraction electrode, 8-grounded electrode, 9- insulator, 11- back flange, 12- gas inlet, 13- view port, 14- cooling water inlet-outlet.

**FIGURE 18**: Simplified version of RF SPS with a solenoidal magnetic field.
1-Gas discharge chamber (AlN), 2- cooling jacket from KEEP, 3- solenoid, 4- saddle antenna, 5-plasma electrode with conical collar and emission aperture, 6-extractor insulator, 7- extraction electrode, 8-permanent magnets, 9-grounded electrode, 10- insulator, 11- back flange, 12- gas inlet, 13- window, 14- cooling water inlet-outlet, 15-shelf, 16- pellets.

For positive ion extraction was prepared simplified version of RF Ion Source with solenoidal magnetic field shown in Fig.18. In this version of Ion Source permanent magnets was removed and extractor electrode was made thinner. In CW mode of operation a discharge can be triggered at high gas pressure and after discharge triggering gas pressure can be decreased. In this case we don't need the triggering plasma gun.

## CW OPERATION

Continuous wave (CW) operation of the SA SPS has been tested in the small SNS test stand. The general design of the CW SA SPS is based on the pulsed version considered above [3, 4]. Some modifications were made to improve cooling and cesiation stability. For CW operation was prepared simplified version of RF SPS with solenoidal magnetic field shown in Fig. 18. In CW mode of operation a discharge can be triggered at high gas pressure and after discharge triggering gas pressure can be decreased. In this case we don't need the triggering plasma gun.

The extracted collector current can be increased significantly by optimization of the longitudinal magnetic field in the discharge chamber. CW operation of the SA SPS with negative ion extraction was tested with RF power up to 2 kW from generator (~ 1.5 kW in the plasma) with production up to Ic=10 mA. Long term operation was tested with 1.5 kW from the RF generator (~ 1 kW in the plasma and 0.5 kW is dissipated in the antenna and matching network) with production of Ic=7 mA, Iex ~15 mA (Uex=8 kV, Uc=14 kV).

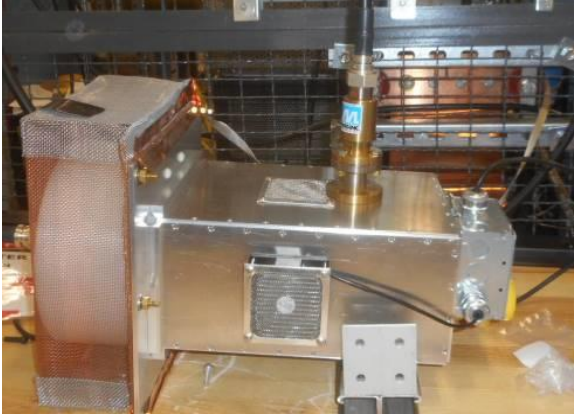
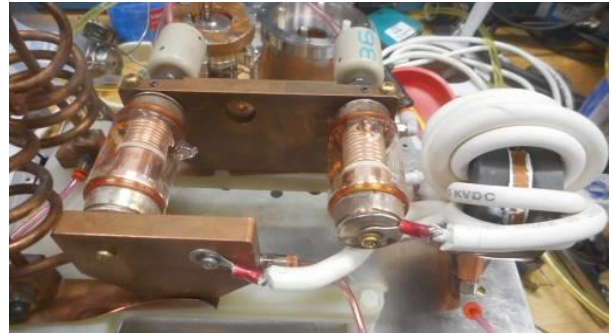

**FIGURE 19.** Photograph of insulating transformer for 13 MHz.

**FIGURE 20**: Photograph of matching network for 13 MHz.

The insulation transformer assemble for 13.56 MHz operation of the SA RF SPS is shown in Fig. 19. Matching network photograph is shown in Fig. 20.

The initial cesiation of the SPS was performed by heating the cesium chromate cartridges by discharge and by further heating in an oven to decompose compounds and alloys. The collector was heated by the beam up to 1000°C. This is good evidence of negative ion beam extraction to the collector. The collector current decreases significantly and the extraction (electron) current increases when cesiation is not optimal. Dependence of collector current Ifc on RF power from RF generator and from discharge power in plasma (upper scale) is shown in Fig. 21.

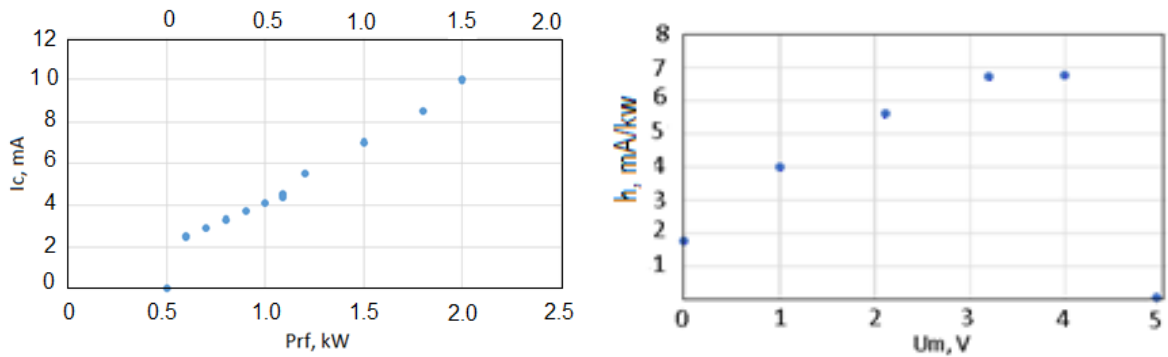

**FIGURE 21:** Dependence of collector current Ifc on RF power from RF generator and from discharge power in plasma (upper scale)

**FIGURE 22:** Dependence of H- ion generation efficiency (in mA/kW) on solenoid voltage Um at 13.56 MHz.

Dependence of H- ion generation efficiency (in mA/kW) on solenoid voltage Um at 13.56 MHz is shown in Fig. 22. The specific power efficiency of negative ion beam production in CW mode is up to Spe = 18 mA/cm$^2$ kW. (In the existing RF SPS the Spe ~ 2-3 mA/cm$^2$ kW [16]; in the TRUIMF filament arc discharge negative ion source [18] the best Spe is about 2 mA/cm$^2$ kW; in a compact Penning discharge SPS [14,19] the Spe is 150 mA/cm$^2$ kW). During H- beam extraction from an external antenna with RF discharges in the AlN discharge chamber, a slow degradation of H- beam intensity after cesiation was observed [20]. It was suspected that this feature of AlN could not be eliminated. Fortunately, after some treatment of the AlN chamber and flanges, the continuous production of H- beam with pulsed current ~ 50 mA (DF ~6%) was supported for 20 days of operation [20]. The cooling of the AlN discharge chamber that was developed is able to remove the heat delivered by discharge with the average RF power up to 3.6 kW.

A pulsed version of this RF SPS with cesiation and triggering plasma Gun, shown in Fig. 23, can be used for charge exchange injection.

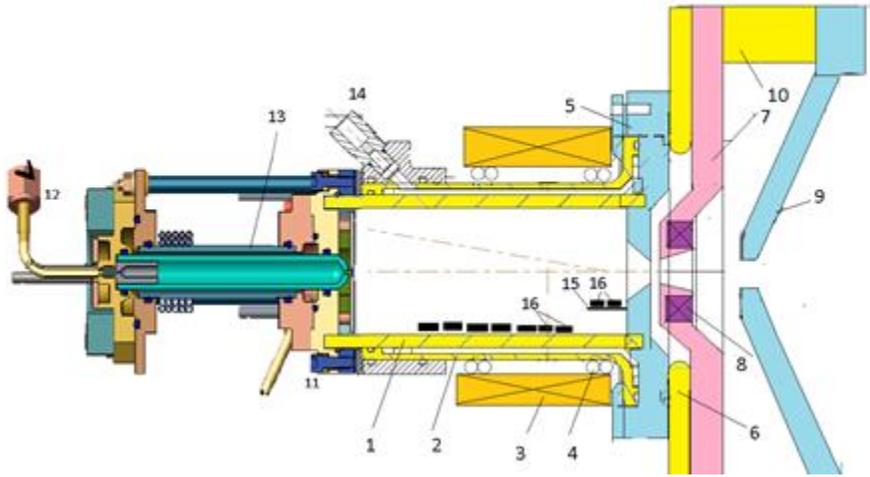

**FIGURE 23**: Pulsed version of simplified RF SPS with cesiation.
1-Gas discharge chamber (AlN), 2- cooling jacket from KEEP, 3- solenoid, 4- saddle antenna, 5-plasma electrode with conical collar and emission aperture, 6-extractor insulator, 7- extraction electrode, 8-permanent magnets, 9- grounded electrode, 10- insulator, 11- back flange, 12- gas inlet, 13- triggering plasma gun, 14- cooling water inlet-outlet from KEEP, 15-shelf, 16- pellets.

## ACKNOWLEDGEMENT

The work was supported in part by US DOE Contract DE-AC05-00OR22725 and by STTR grant, DE-SC0011323.